\documentclass[11pt,twoside]{article}
\usepackage{feynmp}
\usepackage{epsfig}
\newlength{\abstwidth}
\renewcommand{\d}{\ensuremath{\mathrm{d}}}
\renewcommand{\u}{\ensuremath{\mathrm{u}}}
\renewcommand{\b}{\ensuremath{\mathrm{b}}}
\renewcommand{\t}{\ensuremath{\mathrm{t}}} 
\newcommand{\q}{\ensuremath{\mathrm{q}}}
\newcommand{\qbar}{\ensuremath{\bar{\mathrm{q}}}}
\newcommand{\g}{\ensuremath{\mathrm{g}}}
\newcommand{\ti}[1]{\ensuremath{\tilde{#1}}}
\newcommand{\schi}{\ensuremath{\ti{\chi}}}
\newcommand{\sqd}{\ensuremath{\ti{\d}}}
\newcommand{\alphas}{\ensuremath{\alpha_s}}
\newcommand{\squ}{\ensuremath{\ti{\u}}}

\newcommand{\sg}{\ensuremath{\ti{\g}}}
\begin{document}
\begin{flushright}
LU TP 02--35\\
hep-ph/0209199\\
September 2002
\end{flushright}
\begin{center}
{\Large\bf Baryon Number Violation and String Hadronization}\\[10mm]
\end{center}
\hspace*{1cm}\begin{minipage}{\abstwidth}
{\normalsize{\bf P.\ Z.\ Skands} (speaker)} \\
{\normalsize \it Report on work performed in collaboration with T.\
  Sj\"{o}strand.}\\
{\normalsize Theoretical Physics, Lund University, S\"olvegatan 14A,
  S-22362 Lund, Sweden.}\\[2ex]
{\bf Abstract.}
In supersymmetric scenarios with broken $R$-parity, baryon number
violating sparticle decays are possible. We report on 
the development of a framework allowing detailed
studies with special attention given to the hadronization phase. In our model,
implemented in the {\sc Pythia} event generator,
the baryon number violating vertex is associated with the
appearance of a junction in the colour confinement field. This
allows us to tell where to look for the extra (anti)baryon directly
associated with the baryon number violating decay.
\end{minipage}\\[10mm]
\section{Introduction}
In the Minimal Supersymmetric extension of the Standard Model (MSSM),
the standard particle content, extended to two Higgs doublets, is
doubled up by the presence of superpartners to all normal particles. 
The conservation of a multiplicative quantum number called $R$-parity,
defined by 
$R = (-1)^{2S + 3B + L}$, where $S$ is the particle spin, $B$
its baryon number and $L$ its lepton number, is usually assumed,   
since this prevents fast proton decay and has the nice additional 
consequence of making 
the Lightest Supersymmetric Particle (LSP) 
stable, thus making it a WIMP type dark 
matter candidate. 

However, the choice of $R$-parity conservation 
to prevent fast proton decay is not unique, and due to the distinct
differences in collider phenomenology between models with and without
$R$-parity conservation, 
it is of importance to be well prepared for
all possibilities at present as well as future high-energy experiments. 

With $R$-parity conserved, experimental
SUSY signals would consist of jets, leptons and
missing $E_{\perp}$ from escaping neutrinos and LSP's. In scenarios
with baryon number violation (BNV in the following) 
the main decay product is jets, with only few leptons or
neutrinos, and so observability above QCD backgrounds becomes far from
trivial at hadron colliders such as the Tevatron or the LHC. In order
to carry out realistic studies it is therefore necessary to have a
detailed understanding of the properties of both signal and background
events. The prime tool for achieving such an understanding is to
implement the relevant processes in event generators, where simulated
events can be studied with all the analysis methods that could be used
on the real events. 

In this presentation, we concentrate on the possibility that baryon number
may be broken, resulting in BNV sparticle decays. Sparticle production by BNV,
important when the BNV couplings are large and/or the sparticles are heavy, is
not considered here. In the past, BNV has been modelled
\cite{Herwigmodel,Herwigsusy} and studied 
\cite{Herwigstud} in detail in the \textsc{Herwig} framework, with
emphasis on the perturbative aspects of the production process. In
\cite{baryon}, we
present a corresponding implementation in \textsc{Pythia}, summarized here,
where a special effort is dedicated to the non-perturbative aspects, allowing 
us to address the possibility of obtaining a ``smoking-gun''
evidence that a BNV decay has occurred, with questions such as
\textit{Could the presence of a violated baryon number be directly
observed?} and \textit{If so, what strategy should be used?}. 
In addition, many other differences exist between the \textsc{Pythia}
and \textsc{Herwig} physics scenarios, for parton showers and
underlying events, thereby allowing useful cross-checks to be
carried out and uncertainties to be estimated. 
\section{The BNV Scenario}
The most
general superpotential which can be written down for the MSSM includes 4
$R$-parity odd terms:
\begin{equation}
W^{\mathrm{MSSM}}_{\mathrm{RPV}} =
\frac12\lambda_{ijk}\epsilon^{ab}L_a^iL_b^j\bar{E}^k +
\lambda'_{ijk}\epsilon^{ab}L_a^{i}Q_{b}^{j\alpha}\bar{D}_{\alpha}^{k} +
\frac12\lambda''_{ijk}\epsilon^{\alpha_1\alpha_2\alpha_3}
\bar{U}^i_{\alpha_1}\bar{D}^j_{\alpha_2}\bar{D}^k_{\alpha_3}
+ \kappa_{i}L^i_a H_2^a \label{eq:wmssm}
\end{equation}
where $i,j,k$ run over generations, $a,b$ are SU(2)$_L$ isospin indices,
and $\alpha_{(i)}$ runs over colours. 

In a $B$-conserving theory like the SM or the
$R$-conserving MSSM, there is no colour antisymmetric perturbative
interaction term, i.e.\ no term with a colour structure like that of the UDD
term (the 
third term in the above equation). Apart from extreme occurrences, like
knocking two valence quarks out of the same proton in different directions,
by two simultaneous but separate interactions, normal high-energy events
would therefore not fully display the antisymmetric colour structure of
the proton. So what is different about the UDD term
is that it allows the production of three colour carriers at large momentum
separation, without the creation of corresponding anticolour carriers.
It is the necessary SU(3) gauge connection between these three partons
 that will lead us in the development of the nonperturbative framework.
 
A further point about the UDD term is that the contraction of the
$\epsilon$ tensor with $\bar{D}^j\bar{D}^k$ implies that $\lambda''_{ijk}$
should be chosen antisymmetric in its last two indices, since a
$(j,k)$-symmetric part would cancel out.
 
The part of the Lagrangian
coming from the UDD superpotential term in which we are interested is:
\begin{equation}
\mathcal{L}_{\mathrm{BNV}} = {\textstyle\frac12}\lambda''_{ijk}
\epsilon^{\alpha_1\alpha_2\alpha_3}
\left(\bar{u}^i_{R\alpha_1}(\tilde{\d}^*)^{j}_{R\alpha_2}(\d^c)_{R\alpha_3}^k
+\bar{\d}^j_{R\alpha_1}(\tilde{\u}^*)^i_{R\alpha_2}(\d^c)^k_{R\alpha_3}
- (j\leftrightarrow k)
\right)+ h.c. \label{eq:bnvlag}
\end{equation}
where we have made the choice of not yet using any of the antisymmetry
requirements, so that the ordinary Einstein summation convention applies.

Combining the vertices in eq.~(\ref{eq:bnvlag})
with the full MSSM Lagrangian, also decays involving one or
more gauge couplings are clearly possible, e.g.\ neutralino decay via
$\schi^0\to\tilde{\q}_i(\to \bar{\q}_j\bar{\q}_k)\bar{\q}_i$. The BNV
SUSY decay processes currently implemented in \textsc{Pythia}, with Born
level matrix elements as calculated by
\cite{Herwigmodel}, are:\vspace*{1mm}\\
\begin{tabular}{llclr}
1)\hspace*{4mm}&$\sqd_{jn}$&$\to$&$\bar{\u}_i\bar{\d}_k$ & (36) \vspace*{1mm}\\
2)&$\squ_{in}$&$\to$&$\bar{\d}_j\bar{\d}_k$ & \hspace*{5mm}(18) \vspace*{1mm}\\
3)&$\schi^0_n$&$\to$&$\u_i\d_j\d_k$ & (144) \vspace*{1mm}\\
4)&$\schi^+_n$&$\to$&$\u_i\u_j\d_k$ & (30) \vspace*{1mm}\\
5)&$\schi^+_n$&$\to$&$\bar{\d}_i\bar{\d}_j\bar{\d}_k$ & (14) \vspace*{1mm}\\
6)&$\sg$&$\to$&$\u_i\d_j\d_k$ & (36) \vspace*{1mm}\\
\end{tabular}\\
where $n$ runs over the relevant mass eigenstates: $n\in\{L,R\}$ for the
 first two generations of squarks, $n\in\{1,2\}$ for the third generation
 squarks and the charginos, and
 $n\in\{1,...,4\}$ for the neutralinos. The numbers in
brackets are the number
of modes when summed over $n$, $i$, $j$, and $k$, and over charge conjugate
modes for the Majorana particles.
 
When calculating the partial widths (and hence also the rates)
into these channels, we integrate these matrix elements over the
full phase space with massive $\b$ and $\t$ quarks, massive $\tau$
leptons, and massive sparticles. All other particles are only treated as
massive when checking whether the decay is kinematically allowed or not.
 
A feature common to the \textsc{Herwig} and \textsc{Pythia}
implementations is how double-counting in the BNV three-body
modes is avoided. The diagrams for these modes contain intermediate squarks
which may be either on or off the mass shell, depending on the other masses
involved in the process. If a resonance can be on shell, we risk doing double
counting since \textsc{Pythia} is then already allowing the process, in the
guise of two sequential $1\to 2$ splittings. In particular, this means that
the list of $1\to3$ BNV widths obtained by a call to
\texttt{PYSTAT(2)} only represent the non-resonant contributions, the resonant
ones being accounted for by sequences of $1\to 2$ splittings in
other parts of the code.
                                                                            
\section{BNV Colour Topologies}
Up till now we have considered short-distance processes, where
perturbation theory provides a valid description in terms of quarks,
gluons and other fundamental particles. At longer distances, the
running of the strong coupling $\alphas$ leads to confinement and a
breakdown of the perturbative description of QCD processes. 
The perhaps most successful
and frequently model for the transition from the description in terms of 
quarks and gluons to a description based on hadrons is 
the Lund string fragmentation model \cite{Lundstring}.
 
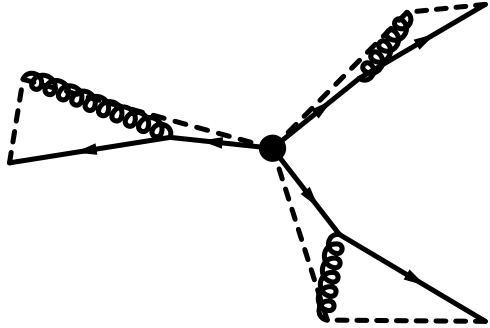
\begin{figure}\vspace*{-4mm}
\center\setlength{\unitlength}{0.5pt}
\begin{fmffile}{fmfbvjup}
\begin{fmfgraph*}(400,240)\fmfset{curly_len}{2mm}\fmfset{arrow_len}{3mm}
\fmfleft{l0,l1,l2,l3,l4}\fmfpen{thin}
\fmftop{t0,t1,t2,t3,t4}
\fmfbottom{bm2,bm1,b0,b1,b2,b3}
\fmfright{r1,r2,r3}
\fmf{fermion}{v1,r1}
\fmf{fermion,tension=2}{v,v1}
\fmf{gluon}{v1,b1}
\fmf{fermion}{v3,r3}
\fmf{fermion,tension=2}{v,v3}
\fmf{gluon}{v3,t3}
\fmf{fermion,tension=3}{v,v2}
\fmf{fermion}{v2,l2}
\fmf{gluon}{v2,l3}
\fmfv{d.sh=circ,d.siz=0.07h}{v}
\fmffreeze
\fmfpen{thick}
\fmf{dashes}{l2,l3,v}
\fmf{dashes}{r3,t3,v}
\fmf{dashes}{r1,b1,v}
\end{fmfgraph*}
\end{fmffile}
\caption{String drawing in a BNV colour topology. The full lines represent
  quarks going out from the decay vertex, the curly lines gluons emitted in
  the parton shower, and the dashed lines the final strings stretched from
  each quark across its colour connected gluons back to the junction. Note:
  this picture was drawn in a ``pedagogical projection'' where distances
  close to the center are greatly exaggerated.\label{fig:junction}}
\end{figure}
This approach has not before been applied to the colour topologies
encountered in BNV. Therefore we here extend the model by the
introduction of a junction, where three string pieces come together, c.f.\
figure \ref{fig:junction}.
Effectively, it is this junction that carries the (anti)baryon number
that is generated by a BNV process. The hadronization in the region
around the junction will therefore be of special interest.

In figure \ref{fig:junction}, 
the central black dot represents such a junction, and the dashed lines show
the string pieces stretched between the junction and each 
endpoint quark, across emitted gluons, resulting in a Y-shaped topology. 
In the simplest picture of
fragmentation, each string piece is broken by the formation of a number of
\q\qbar\ pairs along the string. The end-point quark of each piece then pairs
up with the closest \qbar\ (in colour space) to form a meson, leaving a new unpaired \q\ which
pairs up with another \qbar, and so on until almost 
all the energy stored in each
string piece is used up. From this picture, it is evident that the
fragmentation eventually produces 3 unpaired quarks, one on each side of 
the junction. By colour conservation, with the split off mesons being colour
singlets, these 3 quarks are in a colour-antisymmetric state, i.e.\ a
baryon. In the following, we refer to this baryon as the ``junction baryon''. 

It could have been interesting to contrast the junction concept
with some alternatives, but we have been unable to conceive of any
realistic such, at least within a stringlike scenario of confinement.
The closest we come is a V-shape topology, with two string pieces,
similar to the configuration in a $\q\qbar\g$ topology. This
would be obtained if
one e.g. imagined splitting the colour (anti-colour) of one of the final
state quarks (antiquarks)
into two anticolours (colours). In such
a scenario the baryon would be produced around this quark, and could
be quite high-momentum. Of course, such a procedure is arbitrary,
since one could equally well pick either of the three quarks to be in
the privileged position of producing the key baryon. Further,
with two string pieces now being pulled out from one of the quarks,
the net energy stored in the string at a given (early) time is
larger than in the junction case, meaning the Y junction is
energetically favoured over the V topology. For these reasons, the 
V scenario has not been pursued.   

\subsection{Fragmentation of Junction Strings}
As mentioned, the kind of string configuration depicted in
fig.~\ref{fig:junction} has not previously been a part of \textsc{Pythia},
thus we here outline the technical aspects of the fragmentation process step
by step. A more comprehensive description will be contained in \cite{baryon}.

In the rest frame of the junction the opening angle between any pair of
quarks is 120$^{\circ}$, i.e.\ we have a perfect Mercedes topology.
This can be derived from the action of the classical string
\cite{artrujunc}, but follows more directly from symmetry arguments.

Using this requirement, the rest frame of the junction can easily be found
for the case of 
three massless quarks (and no further gluons), but the general massive
case admits no analytical solution. Rather, we use an iterative, numerical
procedure. 

When gluon emission is included, the junction motion need not be uniform.
Consider e.g.\ an event like the one in fig.~\ref{fig:junction}.
Here the quarks each radiated a gluon, and so the strings
to the junction are drawn via the respective gluons. 
It is the direction of these gluons that determines the junction motion at
early times, and the directions of the quarks themselves are irrelevant.
As a gluon moves out from the junction origin, it loses energy to the string. 
From the point when it has lost \emph{all} its energy and onwards, it would
then  
be the direction of the respective quark, and not of the gluon, that
defines the pull on the junction, resulting in a ``jittering around'' of the
junction. Naturally, this also applies in the general case where an arbitrary
number of gluons is emitted.

Rather than trying to trace this jitter in detail --- which
anyway will be at or below the limit of what it is quantum mechanically
meaningful to speak about --- we define an effective pull of each string
on the junction as if from a single particle with a four-momentum
\begin{equation}
p_{\mathrm{pull}} = \sum_{i=1}^n \, p_i \, \exp \left( -
{\textstyle \sum}_{j=1}^{i-1}
E_j / E_{\mathrm{norm}} \right) ~.
\label{eq:ppull}
\end{equation}
Here $i=1$ is the innermost gluon, $i=2$ is the next-innermost one, and
so on up to $i=n$, the endpoint quark. The energy sum in the exponent runs
over all gluons inside the one considered (meaning it vanishes for $i=1$),
and is normalized to a free parameter $E_{\mathrm{norm}}$, which by default
we associate with the characteristic energy stored in the string at the time
of breaking. Note that the
energies $E_j$ depend on the choice of frame. A priori, it is the energies in
the rest frame of the junction which should be used in this sum, yet since
these are not known to begin with, we employ an iterative procedure.

Since the string junction is a very localized part of the full string system,
it is not desireable that the hard part of the fragmentation spectrum of each
string, i.e.\ the hadrons produced close to the endpoint quark, should be
significantly affected by the presence of the junction.
In particular, if we consider events where each of the three outgoing
quark jets have large energies in the junction rest frame, the production
of high-momentum particles inside a jet should agree with the one of a
corresponding jet in an ordinary two-jet event. This can be ensured by
performing the fragmentation from the outer end of the strings inwards,
just like for the ordinary $\q\qbar$ string. Thus an iterative 
procedure can be used, whereby the leading $\q$ 
is combined with a newly produced $\qbar_1$,
to form a meson and leave behind a remainder-jet $\q_1$, which is fragmented
in its turn. Flavour rules, fragmentation functions and handling of
gluon-emission-induced kinks on the string are identical with the ones
of the ordinary string.
 
While these hadronization principles as such are clear, and give the bulk
of the physics, there is a catch: if all three strings are
fragmented until only little energy and momentum remain in each, and then
these remainders are combined to a central baryon, what guarantees that
this baryon obtains the correct invariant mass it should have?
 
In this brief summary, we are forced to  refer the reader to 
\cite{baryon} for the technical details pertaining to the answer to this
question. The end result is that a physical mass for the junction baryon 
is obtained by first fragmenting two 
of the three strings from the respective end inwards,
towards a fictitious other end. In order to have a
large-mass system left for the system in which energy-momentum conservation 
will eventually be imposed as a constraint, we
prefer to pick these two to be the ones with lowest energy, as defined
in the junction rest frame. As hadrons are successively
produced in the fragmentation, 
their summed energy (in the same frame) is updated. 
Once the hadronic energy exceeds the string energy, 
the fragmentation has gone too far, i.e.\ it has passed the junction point of
the string system, so it is stopped and the latest hadron is rejected.

When two acceptable hadronic chains have been found, the remaining
four-momenta from the respective two strings are combined into a single
parton (diquark), which then replaces the junction as endpoint for the third
string. If the new parton does not turn out to be spacelike, the 
fragmentation procedure for this string is then identical with that of an
ordinary string from here on. Otherwise, the fragmentation is restarted from
the beginning. Note that popcorn 
 baryon production may result in the splitting off of a meson 
from the initial diquark to produce a new diquark. That is, the baryon
number may then migrate to higher energies than otherwise, but will still
be rather centrally produced.

At this point, 
it is interesting to see how dependent our model is on the implicit
assumptions that go into it, for example the definition of the junction pull
vector, eq.~(\ref{eq:ppull}), and the choice of the two least energetic string
pieces as the ones to be fragmented first in the fragmentation scheme
described above. 
\begin{figure}
\center\vspace*{-.75cm}
\includegraphics*[scale=0.7]{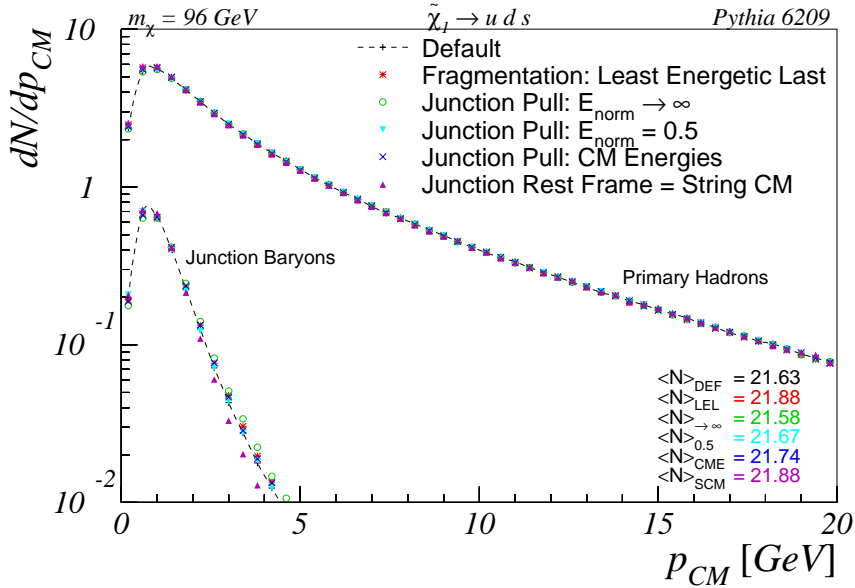}\vspace*{-4mm}
\caption{Momentum spectra of primary hadrons and junction baryons
in the decay of a 96~GeV neutralino to three quarks.
Results with the default implementation are compared with five alternative
ones. $E_{\mathrm{norm}}$ refers to the normalization energy in 
eq.~(\ref{eq:ppull}). Average multiplicities of primary hadrons are shown in
the lower right corner of the plot.
\label{fig:fragdiff}}
\end{figure}

The variation of the CM momentum spectrum of primary hadrons and junction
baryons under changes to these assumptions are shown in figure
\ref{fig:fragdiff} from which it is apparent that the model does not suffer
from stability problems. Observe also that our earlier 
remarks that the junction baryons would be rather centrally produced are
quantified here in the much sharper peaking (notice the log scale) of the
junction baryon momentum
distribution as compared to that of the primary hadrons. With respect to
 the
normalization difference between the two sets of curves, it is 
chiefly due to the many
mesons produced in the fragmentation. The junction baryons in fact roughly 
double the total number of baryons in the momentum region below $\sim$ 2 GeV.
This gives us our
first hint of how to search for this ``smoking-gun'' evidence of BNV.

As a final comment, 
it should be mentioned that more complicated topologies than the ones
so far mentioned are possible. Specifically when two colour-connected BNV
processes occur, there will either 
be two junctions with a string spanned between them or the two baryon numbers
will cancel against each other and give rise to two unconnected $\q\qbar$
string pieces. In the current \textsc{Pythia} implementation, we assume 
that the junction-junction string topology dominates over the non-junction
one, essentially since we expect the string length, and hence the total string
energy, to be smaller more often for the former topology than for the latter.

\section{Conclusion}
It has not previously been possible to study baryon number violating decays
of SUSY particles within the \textsc{Pythia} framework, essentially because
it lacked a hadronization mechanism for colour configurations containing
non-zero baryon number. From \textsc{Pythia} 6.207 on this is now possible,
and the various aspects of the 
implementation have been described in broad terms here. 
Details will be available
in \cite{baryon} and in the \textsc{Pythia} manual. The hadronization 
is based on a physical picture and shows negligible model
dependence. Furthermore, it allows us to ``predict'' that the smoking-gun
evidence of baryon number violation, an excess of baryons, should be looked
for in baryons having small momenta relative to their parent sparticle.

\end{document}